# ANALYSIS OF THE $Y(4500)$ AS THE VECTOR MOLECULAR STATE

**Elif Güngör**
Department of Physics, Kocaeli University - 41001 Izmit, Turkey
*elif-gungor3011@hotmail.com*

**Hayriye Sundu**
Department of Physics, Kocaeli University - 41001 Izmit, Turkey
*hayriye.sundu@kocaeli.edu.tr*

**Jale Yılmazkaya Süngü**
Department of Physics, Kocaeli University - 41001 Izmit, Turkey
*jyilmazkaya@kocaeli.edu.tr*

**Elşen Veli Veliev**
Department of Physics, Kocaeli University - 41001 Izmit, Turkey
*elsen@kocaeli.edu.tr*

**Abstract**

The study of hadronic parameters of mesons is of great importance for understanding the nature of strong interactions and QCD vacuum. One of the approaches to investigate these properties is the QCD Sum Rule (QCDSR) Method, which has been extensively used for studying masses, decay constants, form factors, decay widths, and so on properties of both traditional and exotic hadrons. Recently BESIII collaboration discovered a new particle with a statistical significance greater than $8\sigma$ in the hidden-charm channel of the $e^+e^-$ annihilation, denoted as $Y(4500)$. In this study, we investigate the $Y(4500)$ state in the framework of the QCDSR approach, assuming this particle as a molecular state with quantum numbers $J^{PC} = 1^{--}$. We examined the two-point correlation function and we take into account non-perturbative contributions in OPE side evaluating up to operator dimension 10. We calculate the mass and decay constant of $Y(4500)$ state and obtain a mass value consistent with results in the literature. We hope the studies presented in this paper provide new insights to establish the hidden-charm molecular spectrum.

# 1. INTRODUCTION

The Constituent Quark Model (CQM) proposed by Gell-Mann and Zweig is an effective model for classifying all known hadrons (Gell-Mann, 1964, Zweig, 1964). A conventional meson is a bound state of one quark and one anti-quark, and a baryon is composed of three quarks or three anti-quarks. This representation is remarkably successful in predicting hadron properties and the existence of additional hadrons that were subsequently detected. On the other hand, the question came to the fore as to why hadrons with different quark combinations than ordinary hadrons could not exist. Hence many theoretical and experimental studies in high energy physics have been focused on finding evidence of exotic particles which do not fit the quark model of conventional hadrons. These include tetraquarks (diquarks and anti-diquarks) or meson molecules bound by an attractive interaction, hybrids (quarks, antiquarks, and nonvirtual gluons), glueballs (objects made entirely of gluons), pentaquarks (four quarks plus one antiquark), hexaquarks, resembling two baryons bound together (a dibaryon), or three quarks and three antiquarks (Chen, 2023, Hosaka, 2016).

As the first state in the charmonium family, the $J/\psi$ at Brookhaven and SLAC in 1974 heralded the arrival of a new era in the study of elementary particles, which confirmed the discovery of the charm quark (Aubert, 1974). Since then, a great number of charmonia have been reported by experiments as listed in PDG (Workman, 2022). While the charmonium family has become abundant in experiments, it is far from being well established.

The situation has changed since 2003, as higher energies became available in the experimental facilities giving a chance to hunt the heavy-flavoured multi-quark states (Abazov, 2004). The so-called $XYZ$ states could not be explained with the CQM. In this context, the heavy-light hadrons have been got much attention theoretically and experimentally in the last decade (Agaev, 2022a, Agaev, 2017, Agaev, 2020, Brambilla, 2020). Notably, with the observations of various charmonium and bottonium-like $XYZ$ states, the study of multi-quark hadrons has entered a new era. The charmonium and bottonium spectrum provide a much cleaner environment because of the masses of the heavy quarks ($c$ and $b$) are large and well separated from the QCD scale. So, one might hope that non-conventional states containing $c\bar{c}$ and $b\bar{b}$ pairs would be easier to identify and almost free from interference patterns (Sirunyan, 2018, Albaladejo, 2020, Ablikim, 2020).

The $XYZ$ states do not fit with the CQM. Therefore one cannot find proper positions for them in the meson spectrum. Also, some of the charged mesonic states (Liu, 2008) need the presence of an additional $q\bar{q}$ pair. The important thing is that the existence of these exotic particles does not contradict with the QCD theory. So, many theoretical predictions have been made for the exotic states (Brambilla, 2020, Süngü et al., 2022, Süngü et al., 2019a, Süngü et al., 2019b). Since their masses are very close to the meson-meson thresholds, many authors interpret some exotics as possible molecular meson candidates. Some charmed mesons have also been predicted as molecular states and have a rich spectrum (Guo, 2018). For example, $X(3872)$, as it was the first charmonium state that did not reconcile a $c\bar{c}$ picture observed by the Belle Collaboration, did not match a $c\bar{c}$ picture and activated the search for alternative explanations (Abe, 2004).

It has been proposed that $X(3872)$ could be a $D^*\bar{D}$ molecular state (Voloshin, 2006, Sundu, 2016), $D^*_{s0}(2317)$ as a $D^*K$ molecule (Liu, 2022), $T^+_{cc}(3875)$ be a $D^0 D^{*+}$ (Agaev et al., 2022b), $Z^+(4430)$ be a $D^*\bar{D}_1$ molecular state (Liu, 2008), X(5568) as a $B\bar{K}$ molecule (Agaev et al., 2016), $Y(4260)$ is described as predominantly a $D_1\bar{D}$ molecule (Anwar, 2021). The research on molecular states could deepen our understanding of quark-gluon interactions. Thus, the quantitative descriptions of molecular states' properties, such as mass, decay constant, etc., are needed for well comprehending their structures.

Recently, new resonance in hidden-charmed sector referred to as the $Y(4500)$ state decaying into $K^+K^-J/\psi$ has been reported in the BESIII Collaboration (Ablikim, 2022). Its mass is determined to be $M = (4484.7 \pm$ 13.3 ± 24.1) MeV. The resonance is quite narrow, and its width is measured to be Γ = (111.1 ± 30.1 ± 15.2) MeV. The extreme closeness of the newly detected resonance to the meson pair threshold $D_s\bar{D}_{s1}$ is very suggestive of an interpretation of this particle as a 'molecule' dominantly consisting of neutral charmed-strange meson pairs (Dong, 2021, Peng, 2022). The idea of being the exotic molecular state of $Y(4500)$ was investigated in the context of QCDSR model with quantum number $J^{PC} = 1^{--}$. This study will show us whether the $Y(4500)$ prefer to be $c\bar{c}s\bar{s}$ molecular structure or not.

The manuscript is structured as follows. Section 2 presents the QCDSR model used in calculations. In Section 3, we construct the interpolating current, calculate the two-point correlation functions and spectral densities for these interpolating currents and extract the mass and decay constant of $D_s\bar{D}_{s1}$ molecule state. Section 4 discuss the results. The last section is a summary.

## 2. QCD SUM RULES FOR MOLECULE ASSIGNMENT

Hadrons are produced in a scale of energy where the perturbation theory loses its status; hence to explore the physics of hadrons, a non-perturbative approach is required. The QCDSR approach is one of the most well-established non-perturbative techniques applied extensively to hadron physics and has two main features (Shifman et al., 1979);

- It is based on the QCD Lagrangian and
- Free of model-dependent parameters.

QCDSR method employs momentum-space correlation functions of local composite operators (See Eq. (1)). Each operator (i.e., interpolating fields) has been designed using quark and/or gluon fields to have the same quantum numbers of the particle we are dealing with. The fundamental idea of QCDSR is to equate a theoretical description (i.e., OPE or QCD side) of an appropriate correlation function with a physical (i.e., phenomenological or hadronic side) one. So, a QCDSR calculation has two main parts:

**a)** ***OPE Side***: factorization of the short-distance, perturbatively calculable amplitudes, with the long-distance information accumulated by the quark and gluon condensates proposed by Wilson in 1969.

**b)** ***Physical Side:*** a description of the same correlator in terms of hadronic intermediate states through dispersion relation (i.e., Lehmann representation) that puts together a simple ansatz for the spectral density.

Then we match these two definitions, and this procedure is called "quark-hadron duality". Selecting and equating the same structure calculated in both approaches of the correlator, a relation between QCD degrees of freedom and hadronic parameters are obtained. Then applying Borel transformation concerning the momentum of meson, we automatically suppress the contribution of the continuity. In the end, by doing a numerical analysis, we have the physical parameters of the hadron.

Let's begin the calculation now. The mass and decay constant of the $Y(4500)$ resonance are studied by interpreting it as a $D_s\bar{D}_{s1}$ molecule in the framework of QCDSR, which starts with the two-point correlation function [for brevity, in formulas, we use $Y \equiv Y(4500)$]:

$$\Pi_{\mu\nu}(p) = i \int d^4x e^{ip\cdot x} \langle 0|T\{J_\mu(x)J_\nu^\dagger(0)\}|0\rangle, \qquad (1)$$

where $T$ is the time-ordered covariant operator acting on vector current belonging to $Y(4500)$ state, represented with the following interpolating current (Chen, 2022):

$$J_\mu(x) = \frac{i}{\sqrt{2}}\left(\bar{c}_a(x)\gamma_\mu\gamma_5 s_a(x)\bar{s}_b(x)\gamma_5 c_b(x) - \bar{s}_a(x)\gamma_\mu\gamma_5 c_a(x)\bar{c}_b(x)\gamma_5 s_b(x)\right), \quad (2)$$

here $a$ and $b$ are color indices, $c$ and $s$ denote charm and strange quark fields, respectively.

**a)** Let's determine the physical side first:

$$\Pi_{\mu\nu}^{Phys.}(p) = \frac{\langle 0|J_\mu(0)|Y(p)\rangle\langle Y(p)|J_\nu^\dagger(0)|0\rangle}{m_Y^2 - p_Y^2} + \cdots \quad (3)$$

Parametrizing the coupling of the vector meson with $J^{PC} = 1^{--}$, $Y(4500)$, to the current, $J_\mu$, in Eq. (3) in terms of the meson decay constant $f_Y$ as:

$$\langle 0|J_\mu(0)|Y(p)\rangle = f_Y m_Y \varepsilon_\mu \quad (4)$$

and $\varepsilon_\mu$ is the polarization vector of the molecule state $Y$ and can be written as:

$$\varepsilon_\mu \varepsilon_\nu^* = \left(-g_{\mu\nu} + \frac{p_\mu p_\nu}{m_Y^2}\right). \quad (5)$$

By collecting Eqs. (1)-(5), the following expression of the correlator for the physical side is obtained;

$$\Pi_{\mu\nu}^{Phys.}(p) = \frac{m_Y^2 f_Y^2}{m_Y^2 - p^2}\left(-g_{\mu\nu} + \frac{p_\mu p_\nu}{m_Y^2}\right). \quad (6)$$

Then, in order to eliminate the specified subtractions, a Borel transform is applied to Eq. (6). The Borel transform with respect to the momentum $p$,

$$B(Q^2) \to B(M^2) \equiv \lim_{\substack{Q^2, n \to \infty \\ Q^2/n = M^2}} \frac{(Q^2)^{n+1}}{n!}\left(-\frac{d}{dQ^2}\right)^n B(Q^2), \quad (7)$$

introduces instead as a new energy scale, the Borel mass $M$. By doing Borel transformation;

$$B\Pi^{Phys.}(p) = m_Y^2 f_Y^2 e^{-m_Y^2/M^2} \quad (8)$$

and we separate this statement into structures and select a structure including $g_{\mu\nu}$.

**b)** To continue the QCDSR calculation for the $Y(4500)$ meson's mass and decay constant, we need to determine the OPE side. The first step in the calculation of this side is to insert Eq. (2) into the correlator in Eq. (1). After doing some standard manipulations and contracting the quark fields, we reach an expression in terms of light and heavy quark propagator in the correlation function in Eq. (1) which can be written as:

$$\Pi_{\mu\nu}^{OPE}(p) = \frac{i}{2}\int d^4x e^{ip\cdot x}\left\{Tr[\gamma_5 S_c^{bb'}(x)\gamma_5 S_s^{b'b}(-x)]Tr[\gamma_\mu\gamma_5 S_s^{aa'}(x)\gamma_5\gamma_\nu S_c^{a'a}(-x)]\right.$$

$$-Tr[\gamma_5 S_c^{ba'}(x)\gamma_5\gamma_\nu S_s^{a'b}(-x)]Tr[\gamma_\mu\gamma_5 S_s^{ab'}(x)\gamma_5 S_c^{b'a}(-x)]$$

$$-Tr[\gamma_5 S_s^{ba'}(x)\gamma_5\gamma_\nu S_c^{a'b}(-x)]Tr[\gamma_\mu\gamma_5 S_c^{ab'}(x)\gamma_5 S_s^{b'a}(-x)]$$

$$-Tr[\gamma_5 S_s^{bb'}(x)\gamma_5 S_c^{b'b}(-x)]Tr[\gamma_\mu\gamma_5 S_c^{aa'}(x)\gamma_5\gamma_\nu S_s^{a'a}(-x)]\}. \tag{9}$$

Note that one works with the usual light quark propagator in Eq. (9) as follows:

$$S_q^{ij}(x) = i\delta^{ij}\frac{\not{x}}{2\pi^2 x^4} - \delta^{ij}\frac{m_q}{4\pi^2 x^2} - \delta^{ij}\frac{\langle\bar{q}q\rangle}{12} + i\delta^{ij}\frac{\not{x} m_q\langle\bar{q}q\rangle}{48}$$

$$-\delta^{ij}\frac{x^2}{192}\langle\bar{q}g_s\sigma G q\rangle + i\delta^{ij}\frac{x^2 \not{x} m_q}{1152}\langle\bar{q}g_s\sigma G q\rangle - i\frac{g_s G_{\mu\nu}^{ij}}{32\pi^2 x^2}[\not{x}\sigma^{\mu\nu} + \sigma^{\mu\nu}\not{x}]$$

$$-i\delta^{ij}\frac{x^2 \not{x} g_s^2\langle\bar{q}q\rangle^2}{7776} - \delta^{ij}\frac{x^4\langle\bar{q}q\rangle\langle g_s^2 G^2\rangle}{27648} + \cdots, \tag{10}$$

and the heavy quark propagator $S_c^{ij}(x)$ is expressed as:

$$S_c^{ij}(x) = i\int\frac{d^4k}{(2\pi)^4}e^{-ik\cdot x}\left(\frac{\delta_{ij}(\not{k}+m_c)}{(k^2-m_c^2)} - \frac{g_s G_{\mu\nu}^{ij}}{4}\frac{\sigma^{\mu\nu}(\not{k}+m_c)+(\not{k}+m_c)\sigma^{\mu\nu}}{(k^2-m_c^2)^2} + \frac{g_s^2 G^2}{12}\delta_{ij}m_c\right.$$

$$\left.\times\frac{(k^2+m_c\not{k})}{(k^2-m_c^2)^4} + \frac{g_s^3 G^3}{48}\delta_{ij}\frac{(\not{k}+m_c)}{(k^2-m_c^2)^6}[\not{k}(k^2-3m_c^2)+2m_c(2k^2-m_c^2)](\not{k}+m_c)+\cdots\right), \tag{11}$$

here $G_{\mu\nu}^{ij}$ are gluon field strength tensors and can be given as short-hand notation:

$$G_{\mu\nu}^{ij} = G_{\mu\nu}^A t^{ijA}, \quad A = 1,2,3\ldots,8, \tag{12}$$

with $t^{ijA} = \lambda^{ij,A}/2$, being $\lambda^A$ are the Gell-Mann matrices. Applying the Borel transformation on the variable $p^2$ to the invariant amplitude $\Pi^{OPE}(p)$, selecting the same structure in $B\Pi^{Phys.}(p)$, equating the terms corresponding to this structure and subtracting the continuum contribution, one gets OPE side:

$$\Pi^{OPE}(M^2, s_0) = \int_{s_{min}}^{s_0} \rho^{OPE}(s)\, e^{-s/M^2} ds + \Pi(M^2), \tag{13}$$

where $s_{min} = 4(m_c + m_S)^2$ and effective threshold $s_0$ shows the mass of the first excited state with the same quantum number as the selection of interpolating currents for the considered particle. Based on the analyticity, spectral density can be described to the imaginary part of the correlator by dispersion relation as $\rho^{OPE}(s) = \frac{1}{\pi} Im[\Pi^{OPE}(s)]$, and we evaluate up to operator dimension 10:

$$\rho^{OPE}(s) = \rho^{Pert.}(s) + \sum_{N=3}^{8} \rho^{DimN}(s). \tag{14}$$

In Eq.(13), $\Pi(M^2)$ depict some non-perturbative contributions:

$$\Pi(M^2) = \sum_{N=6}^{10} \Pi^{DimN}(M^2). \tag{15}$$

The lengthy expressions of the OPE spectral densities and $\Pi(M^2)$ non-perturbative contributions are not given explicitly in the paper for simplicity.

Finally, taking into account the above expressions, one gets the QCDSR equalities:

$$f_Y^2 = \frac{e^{m_Y^2/M^2}}{m_Y^2} \Pi^{OPE}(M^2, s_0), \tag{16}$$

$$m_Y^2 = \frac{\Pi^{OPE'}(M^2, s_0)}{\Pi^{OPE}(M^2, s_0)}, \tag{17}$$

where $\Pi^{OPE'}(M^2, s_0) = \frac{d\Pi^{OPE}(M^2, s_0)}{d(-1/M^2)}$, and $s_0$ is the cut-off parameter in the vacuum and is not totally arbitrary. The equality (17) can be derived by acting to the expression (16) by the operator $d/d(-1/M^2)$, where $M$ is the auxiliary Borel mass parameter.

## 3. RESULTS

This section presents numerical values of the analytical results for the mass and decay constant of $Y(4500)$ state. The following input parameters for the quark masses $m_c = (1.27 \pm 0.02)$ GeV, $m_s = (93^{+8.6}_{-3.4})$ MeV are taken from Particle Data Group (Workman, 2022), quark condensate $<0|\bar{s}s|0> = -0.8(0.24 \pm 0.01)^3$GeV$^3$ and quark-gluon mix condensates values $\langle\bar{q}g_s\sigma Gq\rangle = m_0^2\langle\bar{q}q\rangle$, and $\langle\bar{s}g_s s\sigma Gs\rangle = m_0^2\langle\bar{s}s\rangle$, and for the gluon condensates $\langle\alpha_s G^2/\pi\rangle = (0.012 \pm 0.004)$GeV$^4$ (Shifman et al., 1979) and $\langle g_s^3 G^3\rangle = (0.57 \pm 0.29)$GeV$^6$ (Narison, 2016) is used with $m_0^2 = (0.8 \pm 0.2)$GeV$^2$ (Dosch et al., 1989).

In the sum rule model, two additional parameters $M^2$ and $s_0$ exist. To extract hadron properties from the QCDSR in Eqs. (16)-(17), one must first choose an acceptable range of Borel mass parameter values, i.e., a Borel interval $(M_{min}, M_{max})$, which will be constrained by some criteria. The lower limit constraint for $M^2$ is controlled by the OPE convergence. The upper boundary for $M^2$ is determined by imposing that the QCD pole contribution should be more than the continuum contribution.

We analyze that our results are reasonably stable as a function of $M^2$ as well. Correlations between the mass and the continuum threshold $s_0$ can be used to avoid inconsistencies in determining these parameters. Ideally, one looks at a minimum in the function $M(s_0)$, which would provide good criteria for fixing both $s_0$ and $M^2$. Doing these standard analyses, this study will take the continuum threshold parameters as $\sqrt{s_0} = (m_Y + 0.15)$ GeV. The comparison between pole and continuum contributions gives:

$$5 \text{ GeV}^2 \leq M^2 \leq 6 \text{ GeV}^2,$$

$$21 \text{ GeV}^2 \leq s_0 \leq 22 \text{ GeV}^2.$$

**Table 1**. Mass and decay constant values of the resonance $Y(4500)$.

|  | Mass (MeV) | Decay constant ($\text{GeV}^4$) |
|---|---|---|
| Our Work | $4488.35 \pm 11.54$ | $(4.04 \pm 0.36) \times 10^{-3}$ |
| Effective Field Theory (Peng, 2022) | $4494.2^{+5}_{-8..94}$ | – |
| Experiment (Ablikim, 2022) | $4484.7 \pm 13.3 \pm 24.1$ | – |

## 4. DISCUSSION

The paper aims to investigate the state Y(4500) using QCDSR theory. This structure's mass is obtained as $m_Y = (4488.35 \pm 11.54)$ MeV, and decay constant is predicted to be $f_Y = (4.04 \pm 0.36) \times 10^{-3}$ $\text{GeV}^4$. Measurement of the decay constant in future experiments may give us some clues as to whether this structure is molecular. We hope the studies presented in this paper provide new insights to establish the exotic hidden-charm molecular spectrum.

## 5. CONCLUSION

This study is the first attempt to evaluate the hadronic parameters of newly discovered resonance $Y(4500)$ in the framework of the QCDSR technique. Our findings are reasonable and consistent with the available experimental data and theoretical studies. The results of the present study may help understand the possible outcomes of future experiments like LHCb. In our plan, it will be of interest to include in the calculations to evaluate the decay width. This work may shed light on the property of the $D_s\bar{D}_{s1}$ interaction and provide qualitative knowledge for future phenomenological research.